\documentclass[10pt,pre,a4paper,twocolumn,superscriptaddress,showpacs]{revtex4-1}
\usepackage[utf8]{inputenc}
\usepackage[colorlinks=true]{hyperref}
\usepackage{amsmath,amssymb}
\usepackage{amsfonts}
\usepackage{graphicx}
\usepackage{color}
\usepackage[dvipsnames]{xcolor}

\begin{document}

\title{Discontinuous shear-thinning in adhesive dispersions}

\author{Ehsan Irani} \affiliation{Berlin Institute for Medical Systems
  Biology, Max Delbrück Center for Molecular Medicine in the Helmholtz
  Association, Berlin, Germany}\affiliation{Berlin Institute of Health
  (BIH), MDC-Berlin, Germany}\affiliation{Institute for Theoretical
  Physics, Georg-August University of G\"ottingen, Friedrich-Hund
  Platz 1, 37077 G\"ottingen, Germany} \author{Pinaki Chaudhuri}
\affiliation{Institute of Mathematical Sciences, Taramani, Chennai 600
  113, Tamil Nadu, India} \author{Claus Heussinger}
\affiliation{Institute for Theoretical Physics, Georg-August
  University of G\"ottingen, Friedrich-Hund Platz 1, 37077
  G\"ottingen, Germany}

\begin{abstract}

  We present simulations for the steady-shear rheology of a model
  adhesive dispersion in the dense regime. We vary the range of the
  attractive inter-particle forces $u$ as well as the strength of the
  dissipation $b$. For large dissipative forces, the rheology is
  governed by the Weisenberg number $ \text{Wi}\sim b\dot\gamma/u$ and
  displays Herschel-Bulkley form $\sigma = \sigma_y+c\text{Wi}^\nu$
  with exponent $\nu=0.45$. Decreasing the strength of dissipation,
  the scaling with $\text{Wi}$ breaks down and inertial effects show
  up. The stress decreases via the Johnson-Samwer law
  $\Delta\sigma\sim T_s^{2/3}$, where temperature $T_s$ is exclusively
  due to shear-induced vibrations. During flow particles prefer to
  rotate around each other such that the dominant velocities are
  directed tangentially to the particle surfaces. This tangential
  channel of energy dissipation and its suppression leads to a
  discontinuity in the flow curve, and an associated discontinuous
  shear thinning transition.  We set up an analogy with frictional
  systems, where the phenomenon of discontinuous shear thickening
  occurs. In both cases, tangential forces, frictional or viscous,
  mediate a transition from one branch of the flowcurve with low
  tangential dissipation to one with larger tangential dissipation.
 
\end{abstract}

\maketitle
\section{Introduction}

Dense dispersions, like colloids or emulsions, display a broad range
of different rheological properties. This variety mirrors the action
of the different forces acting on and between the particles making the
dispersion. In general, it is not at all clear which of these forces
or combinations are relevant for a particular phenomenon on the
continuum level. Still, identification of the relevant players is
needed for a proper design of new materials, which has become
increasingly important in different industrial settings, like food or
cosmetics~\cite{laba93:_rheol_proper_cosmet_toilet}. Using
simulations, simplified model systems can be defined to close this gap
in understanding. By tuning the interaction forces dominant parameter
dependencies can be isolated and the underlying physical mechanisms
identified.

In this contribution, we are interested in the role of different
dissipative forces on flowing adhesive dispersions. The attractive
inter-particle forces in dispersions may be due to various mechanisms,
e.g.  depletion forces~\cite{BecuPRL2006,C3SM27626K} or direct
interactions~\cite{doi:10.1021/la048608z}. In the case of granular
materials, attraction can appear e.g. via the development of capillary
bridges~\cite{HornbakerNature1997,HerminghausAdvPhys2005,MitaraiAdvPhys2006}. % or van der Waals forces [14,15]
At rest, attractive forces quite generally assist in the formation of
clusters or gel-like networks~\cite{kim17:_advan}. In the flowing
state~\cite{PhysRevE.89.050302,bounoua16:_shear} there is then a
continuous competition between the rupturing of the network and
aggregation processes that try to restore local
structure~\cite{MartensSoftMatter2012,CoussotSoftMatter2007}.

\begin{figure}[b]
  \includegraphics[width=0.5\textwidth,keepaspectratio]{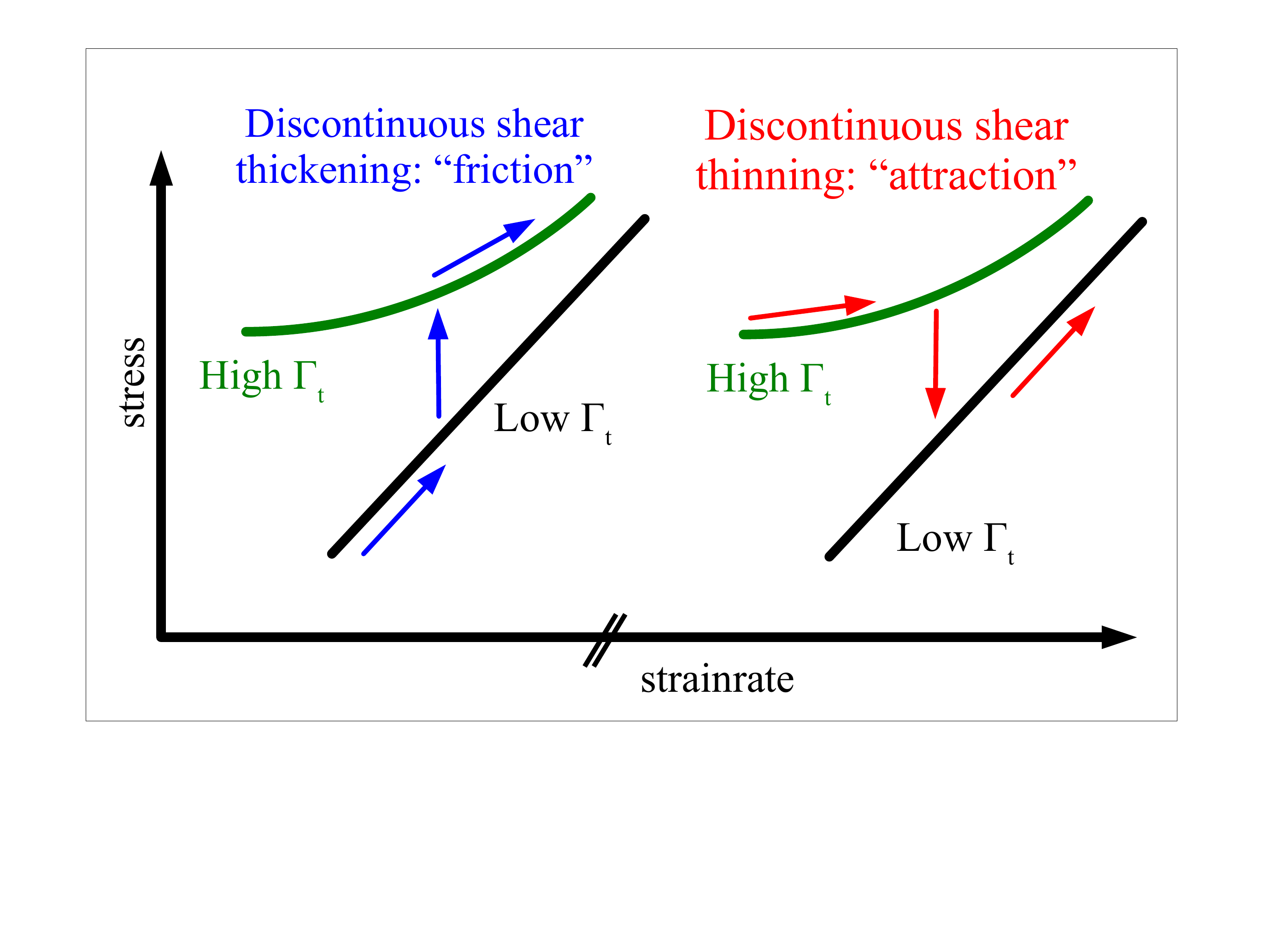}
  \caption{Schematic flowcurves, comparing the two phenomena of
    discontinuous shear-thickening and discontinuous
    shear-thinning. Both are mediated by the presence of a
    ``tangential'' channel of energy dissipation (dissipation is due
    to sliding velocities directed tangentially to particle
    surfaces). Shear thickening is due to a rapid increase of
    frictional dissipation $\Gamma_t$ upon increasing strainrate. On
    the other hand, shear thinning in the adhesive system studied here
    occurs because tangential viscous interactions are only relevant
    in dense gel-like structures, which are broken down upon
    increasing the strainrate.}
  \label{fig:schematic_flowcurve}
\end{figure}

Dissipation in emulsions and suspensions is primarily of hydrodynamic
origin, e.g., in the form of lubrication forces or long-range
hydrodynamic interactions. In granular powders, inelastic collisions
and dry friction dominate the dissipation. In wet granular media,
finally, the breaking of liquid capillary bridges between near by
particles is important~\cite{C2SM25883H}. Many of these forces also
have a directional dependence. Lubrication, for example, has a squeeze
and a shear-mode, the latter often thought to be negligible as to its
logarithmic gap-dependence~\cite{kim05:_microh}. In highly dense
suspensions, close to jamming, it is becoming clear, however, that shear
forces -- acting tangentially to the particle surface -- may
fundamentally affect the emerging rheology. Baumgarten {\it et al.}
~\cite{C7SM01619K} have argued that tangential viscous interactions,
even if they are small, are necessary to obtain dynamic critical
scaling for the linear visco-elasticity of jammed systems. Similarly,
Vagberg {\it et al.}~\cite{PhysRevE.95.052903} demonstrate the key
role that this tangential dissipation plays for the small-strainrate
rheology. In a different system, solid-solid friction between particles
leads to shear-thickening where the associated frictionless system
would only display shear-thinning~\cite{ClausPRE2013}. In particular,
this friction-induced shear-thickening has received a lot of attention
recently~\cite{Clavaud5147,PhysRevX.8.031006,
  PhysRevLett.112.098302,PhysRevE.93.030901}. The
relevant tangential forces in this scenario may even be strong enough
to lead to a discontinuous jump of the stress over several orders of
magnitude, or to a sudden arrest upon increasing the
strainrate~\cite{GrobPRE2014,PhysRevLett.111.218301}. In the granular
community, the combination of frictional interactions with cohesive
forces has been studied in a variety of contexts, e.g. in
Refs.~\cite{SinghPRE2014,SundaresanPRE2014,GilabertPRE2007,BergerEPL2015,KhamsehPRE2015,RognonRouxFLM2008}

In this work, using numerical simulations, we investigate the influence of a tangential viscous force on
the steady-state flow behavior of dense assemblies of adhesive (but non-frictional)
particles. The important finding is that the presence of this additional mode of dissipation
gives rise to a discontinuity in the flow-curves, in the under-damped limit,
quite similar to the phenomenon of discontinuous shear-thickening observed
in frictional systems. In both cases, tangential
forces, frictional or viscous, mediate a transition from one branch of
the flowcurve with low tangential dissipation $\Gamma_t$ to one with
large $\Gamma_t$ (see Fig.~\ref{fig:schematic_flowcurve}). Furthermore, such discontinuities
in the flow curve leads to formation of shear-bands, with contrasting flow rates
and local packing, providing yet another scenario where persistent flow heterogeneities
can happen.
%

%granular community: friction + adhesion motivated by wet granular
%media and liquid bridges, pressure-controlled
%Satoshi/Hayakawa. Rheology of dilute cohesive granular gases~\cite{PhysRevE.97.042902}
%pressure controlled with attraction: \cite{RognonRouxFLM2008,SundaresanPRE2014}

\section{Model}\label{sec:model}

We consider a two-dimensional system of $N$  soft disks interacting via the following potential:
\begin{equation}
\label{eq:pot}
V(r_{ij})=
\begin{cases}
\epsilon\left[(1-\frac{r_{ij}}{d_{ij}})^2-2u^2\right],
& \frac{r_{ij}}{d_{ij}} < 1+u \\
-\epsilon\left[1+2u-\frac{r_{ij}}{d_{ij}}\right]^2,
& 1+u < \frac{r_{ij}}{d_{ij}} < 1+2u \\
0, & \frac{r_{ij}}{d_{ij}} > 1+2u \\
\end{cases}
\end{equation}
where $r_{ij}=|\vec r_i-\vec r_j|$ is the distance between the $i$th
and $j$th particles, and $d_{ij}=(d_i+d_j)/2$ is the average of their
diameters. Thus, there exists a harmonic repulsive interaction when
the particles overlap, $r_{ij}<d_{ij}$. Additionally, there is a
short-range attractive interaction between the particles when the
distance is within some threshold, $d_{ij}<r_{ij}<d_{ij}(1+2u)$.  The
parameter $u$ is introduced to characterize the width ($2u$) and also
the strength ($\epsilon u^2$) of the attractive potential. The scale
for attractive forces is then $\epsilon u/d$. Thus, attractive forces
are characterized by a single parameter. This greatly reduces the
computational complexity and at the same time keeps the model as
simple as possible. In general, we will consider only the case where
the range of attraction is very small as compared to the particle
size. This sets our model apart from LJ-like models, where attraction
usually ranges beyond the first neighbor shell.

In addition to the conservative force, a dissipative force acts
between pairs of particles. This viscous force is proportional to
their relative velocity and acts only when particles overlap, i.e.
$r_{ij} < d_{ij}$,
\begin{equation}\label{eq:fdiss}
	\vec{F}_{\text{diss.}} = -b[\vec{v}_i - \vec{v}_j]
\end{equation}
where $b$ is the damping coefficient. In this model, which is
equivalent to the model coined $\text{CD}$ in Ref.~\cite{VagbergPRL2014},
particle rotations are not accounted for. As discussed in that
reference, rotations are decoupled from the translational degrees of
freedom and therefore can be dropped. It should be noted, that this is
different from the standard Cundall-Strack model~\cite{cundall79} for solid
friction between particles. In that model, the dissipative force is
taken as the relative velocity at the \emph{contact}, which also
involves rotations. Here, it is the relative center-of-mass velocity,
which enters the dissipative force law, Eq.~(\ref{eq:fdiss}).

The damping force can be split in components normal and tangential to
the direction defined by the corresponding contact of the two
particles, $\hat n_{ij} = (\vec r_i-\vec r_j)/r_{ij}$. The normal
contribution, for example, reads
\[\vec F_{\rm diss}^{(n)} = -b[\vec v_{ij}\cdot \hat n_{ij}]\hat n_{ij}\,.\]
In previous work, we have studied the rheological properties in
systems with only this normal
contribution~\cite{IraniPRE2016,IraniPRL2014}. Below we will make
frequent comparison with that work.

To investigate the rheology of such a system of particles, we perform
molecular dynamics simulations using LAMMPS~\cite{LAMMPS}. In order to
avoid crystallization, we choose a 50:50 binary mixture of particles
having two different sizes, with a relative radii of 1.4. The system
is sheared in $\hat{x}$ direction with a strain rate $\dot{\gamma}$
using Lees-Edwards boundary conditions. The volume fraction is
$\phi = \sum_{i=1}^{N} \pi (d_i/2L)^2$, where $L$ is the length of the simulation box.

The unit of energy is $\epsilon$ and the unit of length is the
diameter of the smaller particle type, $d = 1.0$. The unit of time is
hence $d/\sqrt{\epsilon/m}$, where $m=1.0$ is the mass of the
particles. We use $\delta t=0.005$, as timestep for the integration.

\section{Results and Discussion}\label{sec:results}

\begin{figure}
 \includegraphics[width=0.5\textwidth]{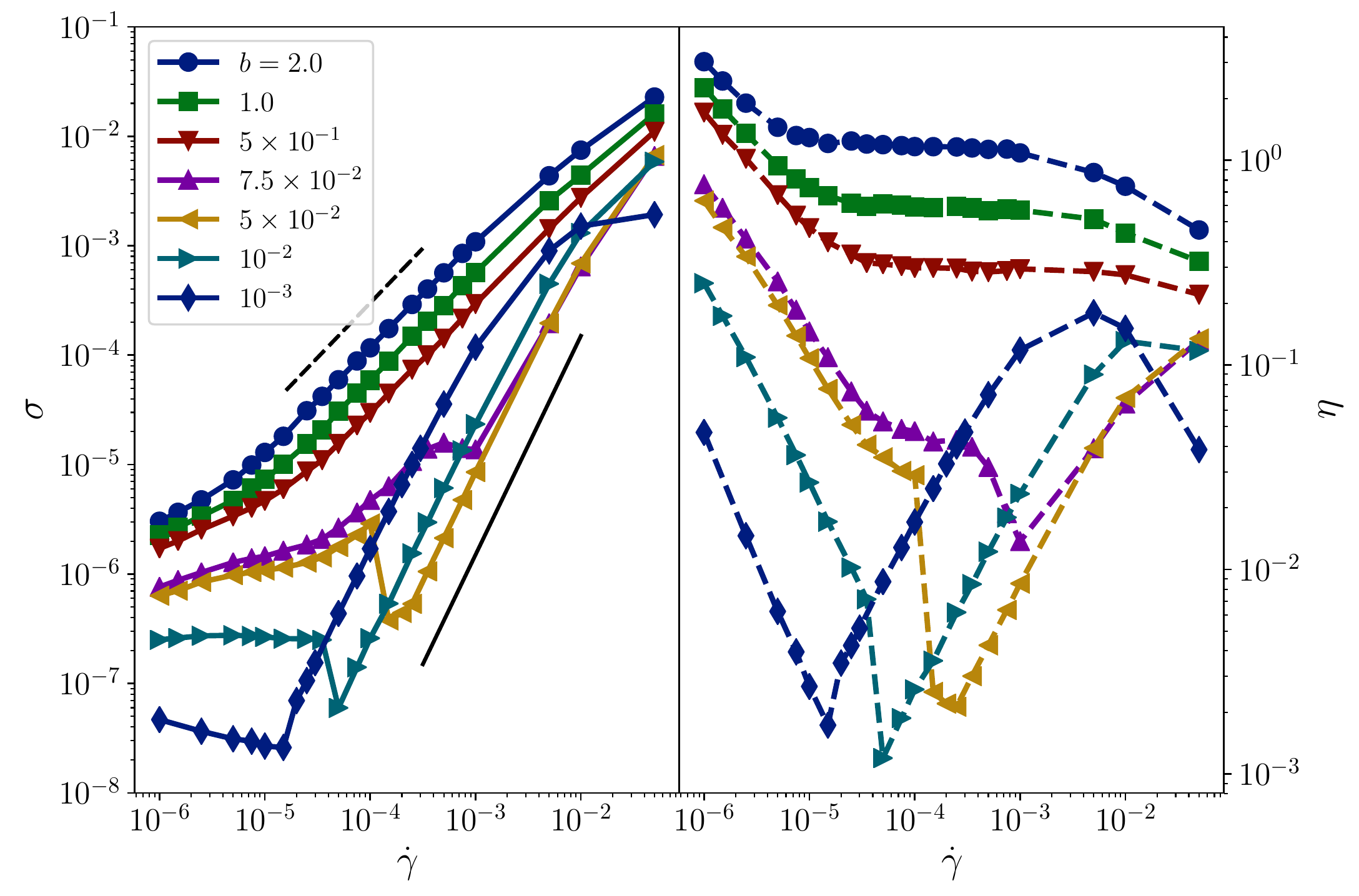}
  \caption{(left) Flow curves (stress $\sigma$ vs shear-rate $\dot{\gamma}$), for
    $u=2\times10^{-5}$ and $\phi=0.75$. As $b$ changes, the system
    crosses from the over-damped to under-damped regime. {Solid
      and dashed lines represent respectivly $\dot\gamma^2$ and
      $\dot\gamma^1$.} (right) Corresponding data for viscosity, $\eta(\dot{\gamma})$.} 
  \label{fig:fc-z-u2e-5}
\end{figure}

For a given imposed strainrate $\dot{\gamma}$ we measure the shear
stress $\sigma$ with the help of the virial expression. The resulting
flowcurves $\sigma(\dot{\gamma})$ are displayed in
Fig.~\ref{fig:fc-z-u2e-5}(a) for different damping parameters $b$ at
$\phi=0.75$. The corresponding viscosity, $\eta(\dot{\gamma})$,
is shown in Fig.~\ref{fig:fc-z-u2e-5}(b).

%At the smallest accesible strainrate $\dot\gamma=10^{-6}$ the various
%curves differ in stress by roughly two orders of magnitude. Still, in
%the limit of vanishing strainrates all these curves have to approach
%the same limiting yield stress $\sigma_y$. In previous work with a
%different damping model \cite{IraniPRE2016,IraniPRL2014} we evidenced
%a finite yield stress for densities as small as $\phi=0.5$. As the
%yield stress is independent of damping, it should be the same in both
%models.

The divergent viscosity for vanishing strainrates indicates the
presence of a yield-stress. For finite strainrates, decreasing the
damping parameter $b$ implements a transition from an overdamped to an
underdamped regime. In the overdamped regime ($b\geq 0.5$), as the
shear-rate is increased, a viscous Newtonian flow regime,
$\sigma\propto \dot\gamma$, is encountered, in the intermediate
$\dot\gamma$ range (dashed line). This translates to a constant
viscosity, before the shear-thinning regime at larger shear-rates.
For the underdamped regime, the situation is very different.  At
larger shear-rates, the stress is $\sigma\propto \dot\gamma^2$ (solid
line), which is called Bagnold regime, wherein the viscosity also
increases linearly with shear-rate. Note the opposite dependence on
$b$ in the two branches. In the viscous branch, the stress decreases
with decreasing damping, while in the case of the Bagnold branch, it
increases.  The latter happens due to the higher velocities in weakly
damped systems.

Recent experiments \cite{FallPRL2010} evidence a simple
continuous transition from viscous to Bagnold scaling. The scenario
encountered here is seemingly more complex 
and even involves discontinuous jumps in the stress
and consequently the viscosity; see Fig.~\ref{fig:fc-z-u2e-5}.

%\bblue{Once the system of particles yields at finite shear-rates,
%there are three particular features that are noteworthy.} The stress
%decreases upon reducing the damping; furthermore, the apparent slope
%of the flowcurve in the small-strainrate regime is reduced and
%eventually changes sign (smallest $b=10^{-3}$). Finally, the
%flowcurves display a discontinuity if the damping is sufficiently
%small, \bblue{with the viscosity abruptly jumping, either decreasing
%or increasing (smallest $b$), from a shear-thinning regime to a
%Bagnoldian shear-thickening regime}.

\subsubsection{Overdamped systems}

In order to understand these unusual features we start by discussing
the overdamped limit, where the Weissenberg number has to be used
to scale the flow curves. The Weissenberg number is defined as the
ratio of dissipative to elastic contributions to the stress. In our
case the dissipative stress scale is given by
$\sigma_{\text{diss}} \sim b\dot\gamma$. The elastic stress scale is
$\sigma_{\text{el.}} \sim \epsilon u/d$. Thus, the Weissenberg number
can be written as
\begin{equation}
  \text{Wi}=\frac{d}{\epsilon}\frac{b\dot\gamma}{u}.
\end{equation}
%in terms of time-scales, $\tau = b/\epsilon u$ is time for ...

\begin{figure}
  \includegraphics[width=0.4\textwidth]{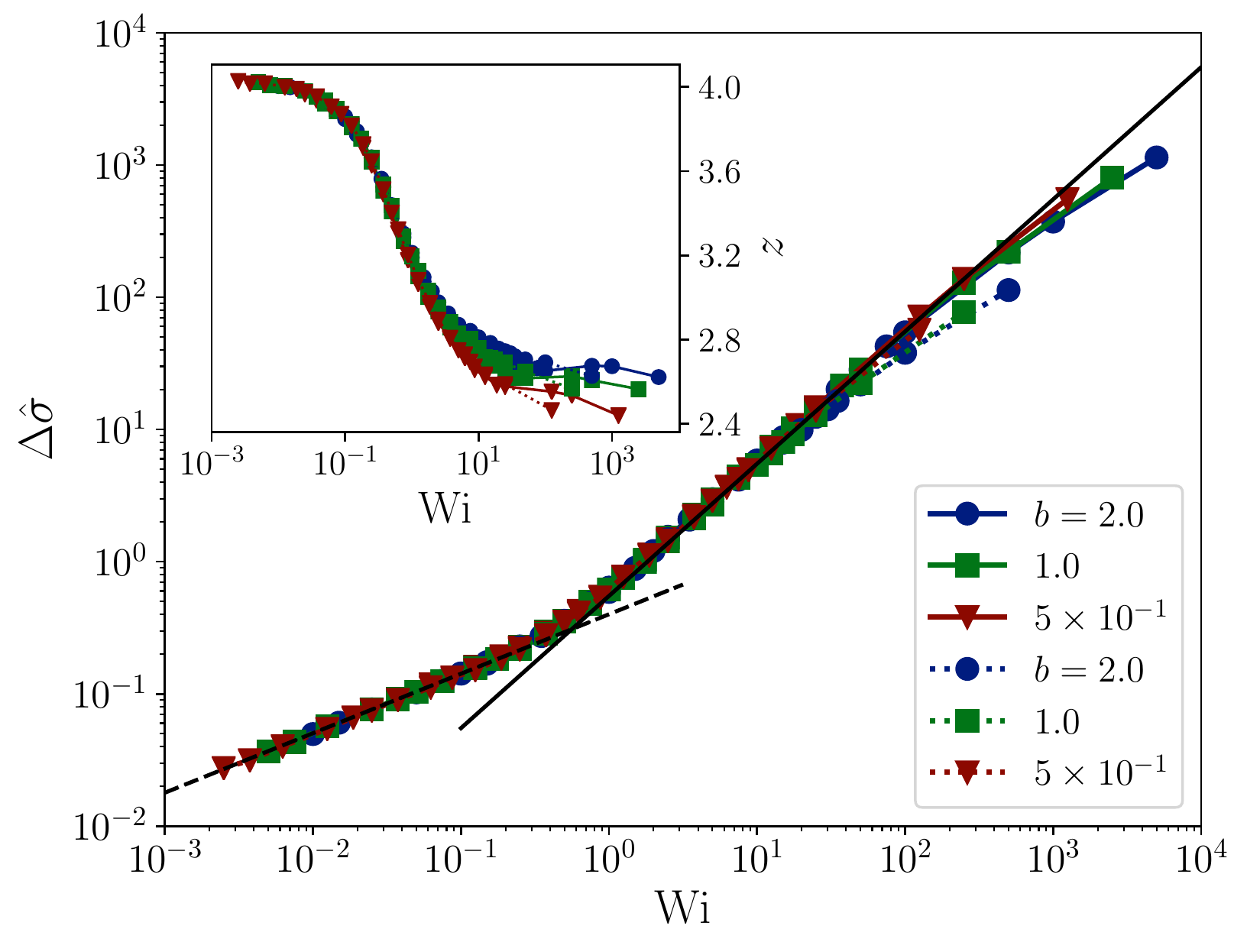}
  \caption{Stress $\Delta\hat\sigma=(\sigma-\sigma_y)/\epsilon u$ vs.
    $\text{Wi}$ of over-damped systems ($b\ge0.5$) at different
    attraction ($u$) and damping ($b$) strength. Solid and dashed
    lines represent, respectively, $u=2\times10^{-5}$ and
    $u=2\times10^{-4}$, with corresponding $\sigma_y=1.8\times10^{-7}$
    and $3.4\times 10^{-6}$. The dashed line without symbols displays
    $\text{Wi}^{\nu}$ ($\nu\approx 0.45$) and the solid line without
    symbols exhibits $\text{Wi}^1$. The inset displays the scaling of
    corresponding connectivity curves $z$ with the Weissenberg number
    ($\text{Wi}$). }
  \label{fig:cd-fc-z-wi}
\end{figure}

The flow stress, ie. after removal of the yield stress, has to be a
function of only Wi,
$\Delta\sigma=\sigma-\sigma_y = \epsilon u F( \text{Wi})$. As the
yield stress itself is not accessible in our simulations we determine
its value so as to enforce this data
collapse. Fig.~\ref{fig:cd-fc-z-wi} displays this collapse for the
stress $\Delta\sigma$ in the overdamped regime for two different
attraction strengths ($u=2\times10^{-5}$ and $u=2\times10^{-4}$)
~\footnote{The values obtained for the yield-stress are
    $\sigma_y(u=2e-5)= 1.8e-7$ and $\sigma_y(u=2e-4)= 3.4e-6$. These
    correspond well with our previous simulations with a different
    damping model~\cite{IraniPRE2016}.}.
Apparently, $ \text{Wi}$ separates two flow regimes, each of which is
characterized by a specific exponent $\sigma \sim \text{Wi}^\nu$.  For
high $\text{Wi}>0.3$ the exponent $\nu=1$, thus
$\sigma \sim b\dot\gamma$ which corresponds to the simple viscous
Newtonian regime mentioned above. The attractive forces in this regime
are irrelevant.

For low $\text{Wi}<0.3$ the exponent $\nu\approx 0.45$. As we are
dealing with a yield-stress fluid this exponent has to be interpreted
as a Herschel-Bulkley (HB) exponent. Thus, the expression for the
stress becomes
\begin{equation}\sigma = \sigma_y + c\frac{\epsilon u}{d}\text{Wi}^{\nu}\,.
\end{equation}
with the yield-stress $\sigma_y$ and a constant $c\approx 0.41$. The
value of the exponent $\nu$ is in the range observed in other
yield-stress fluids (usually between 0.4-0.5), but hardly ever a real
power-law regime (straight line in double-logarithmic presentation) is
observed. In experiments (and many simulations) only small stress
increase over the yield-stress is present.

The yield-stress itself naturally does not scale with the Weisenberg
number and is controlled by different physical mechanisms. These
mechanisms of flow at the yield stress are by now well
understood. Flow comes about as a sequence of elastic branches, during
which the solid is elastically strained, and sudden plastic events, at
wich this stored energy is released and
dissipated~\cite{MaloneyPRE2006}. This gives the stress-strain
relation at the yield stress the typical sawtooth
apearance~\cite{MaloneyPRE2006}. In the elastic branches the stress
increases linearly, $\sigma = g\gamma$, with the slope given by the
elastic shear modulus $g$. The plastic events are quasi instantaneous
if the strainrate is infinitesimal small. The yield-stress can thus be
written in terms of a yield strain $\gamma_y$, as
$\sigma_y=g\gamma_y$.  In previous work
\cite{IraniPRE2016,IraniPRL2014} we have derived the expressions
$g\sim \epsilon\delta\! z_0$ and
$\gamma_y\sim (\delta\! z_0 u)^{1/2}$, relating both quantities to the
connectivity $\delta\! z_0=z(\dot\gamma=0)-4$. The value
$z_{\rm iso}=4$ represents a limiting minimal (isostatic) connectivity
that is necessary for a solid (in 2d) to exist. The value of
$\delta\!z_0$ is generally very small (see inset
Fig.~\ref{fig:cd-fc-z-wi}) indicating the formation of a fragile
solid. It is well known that in these near-isostatic systems
($\delta z\to 0$) the linear response to deformation is characterized
by strong non-affine motion, with particle displacements that are
directed tangentially to particle
contacts~\cite{vanHeckeJPCM2010}. Thus contacting particles tend to
rotate around each other. It is this behavior that leads to the
particular scaling of the yield strain $\gamma_y$. At finite
strainrates displacements translate into velocities, such that the
dominant contribution is a relative velocity $v_{ij}^{(t)}$ of two
contacting particles ($ij$) in direction tangential (${t}$) to the
particle surface. According to the dissipative force law
Eq.~(\ref{eq:fdiss}), this motion gives rise to dissipation which
shows up in the stress via an energy balance equation
\begin{equation}\label{eq:en_balance}
  \sigma\dot\gamma \sim bv^2\,.
\end{equation}
where the left hand side gives the injected work per unit of time,
while the right gives the dissipated power in terms of the
mentioned tangential velocities
\begin{equation}\label{eq:v_tangential}
  v = \sqrt{\langle v_{ij}^{(t)2}\rangle}\,.
\end{equation}
In principle, the velocity component directed normal to the particle surface
also has to be accounted for. But, as will be discussed below, this
latter contribution is very small in the HB regime and thus can be
neglected (see Fig.~\ref{fig:fc-underdamped-repatt-gamma} right).

\begin{figure}[h]
  \includegraphics[width=0.49\textwidth]{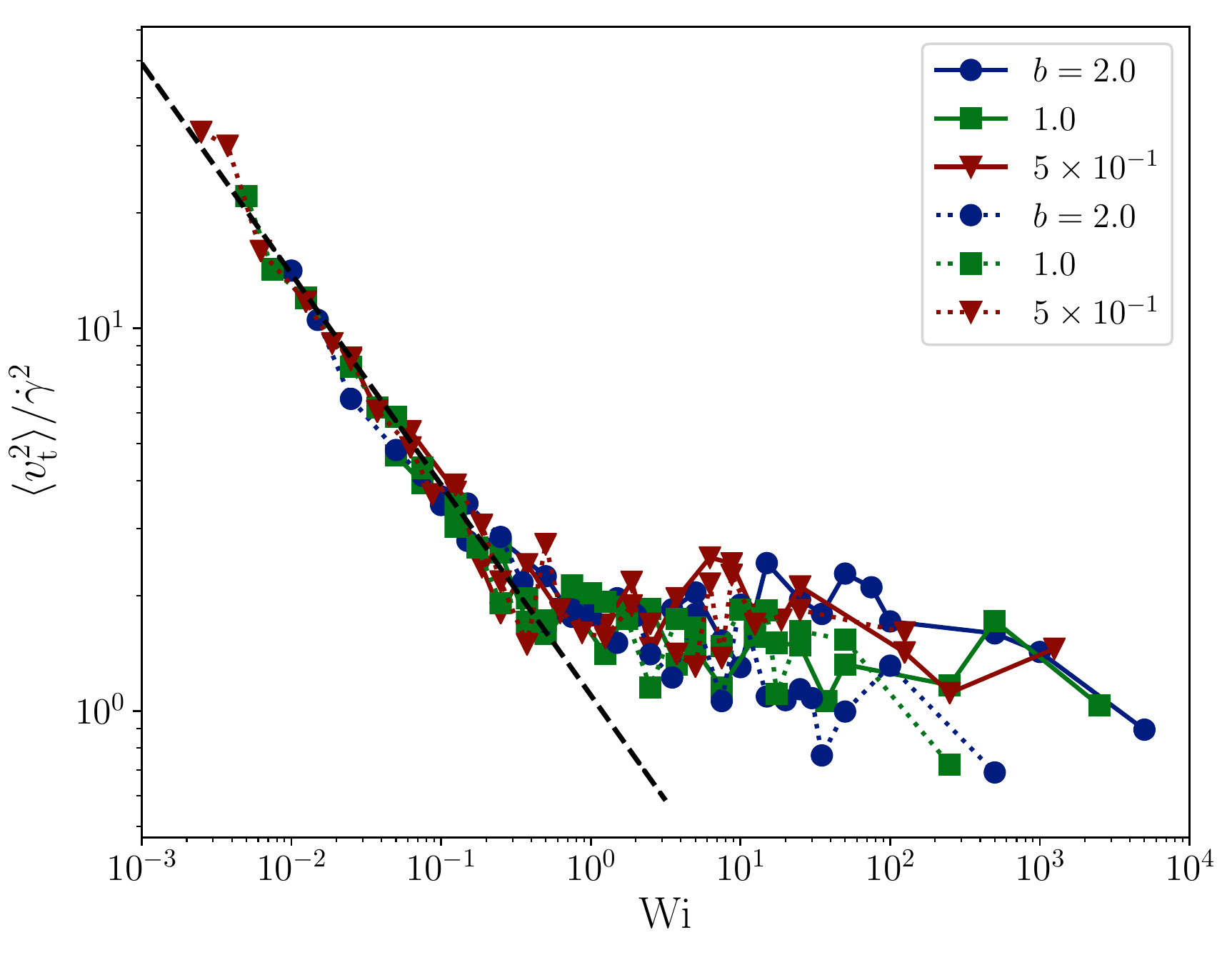}
  \caption{(Particle (tangential) velocities $v^2/\dot\gamma^2$
    vs. $\text{Wi}$ in the overdamped regime for different attraction
    ($u$) and damping ($b\geq 0.5$) strength. For small strainrates
    the velocities scale as $v^2\sim \dot\gamma^2\text{Wi}^{\nu-1}$
    with $\nu-1\approx -0.55$ (dashed line). At large strainrates
    (viscous regime) velocities are nearly independent of $\text{Wi}$
    and $v\sim\dot\gamma$.
  }
  \label{fig:velocities}
\end{figure}

Rewriting Eq.~(\ref{eq:en_balance}) in terms of $\text{Wi}$ and
inserting the HB scaling we get
\begin{equation}\label{eq:v_scaling}
  v^2\sim\dot\gamma^2\text{Wi}^{\nu-1}
\end{equation}
This scaling is tested in Fig.~\ref{fig:velocities}.

\subsubsection{Underdamped systems}

We now turn to the underdamped regime ($b<0.5$). As is evident from
Fig.~\ref{fig:fc-z-u2e-5} the flowcurves in this regime acquire a
different shape than in the overdamped regime. Furthermore, a
discontinuity shows up. The Weisenberg number is therefore no longer
sufficient to explain the flow behavior.

To rationalize these findings we follow Nicolas {\it et al.}
\cite{NicolasPRL2016} and assume that in underdamped systems enhanced
velocities act like a temperature, $T\sim mv^2$, that weakens the solid-like
structure encountered at small strainrates. Weakening comes about
because of activated events that allow plastic rearrangements to take
place even though the threshold of the event is not yet reached. As a
consequence, the overall stress is lowered by a factor
$\Delta\sigma_{\rm therm}$ that we now determine (following
Refs.~\cite{PhysRevLett.95.195501,PhysRevLett.105.266001}).

\begin{figure}[ht!]
  \includegraphics[width=0.45\textwidth]{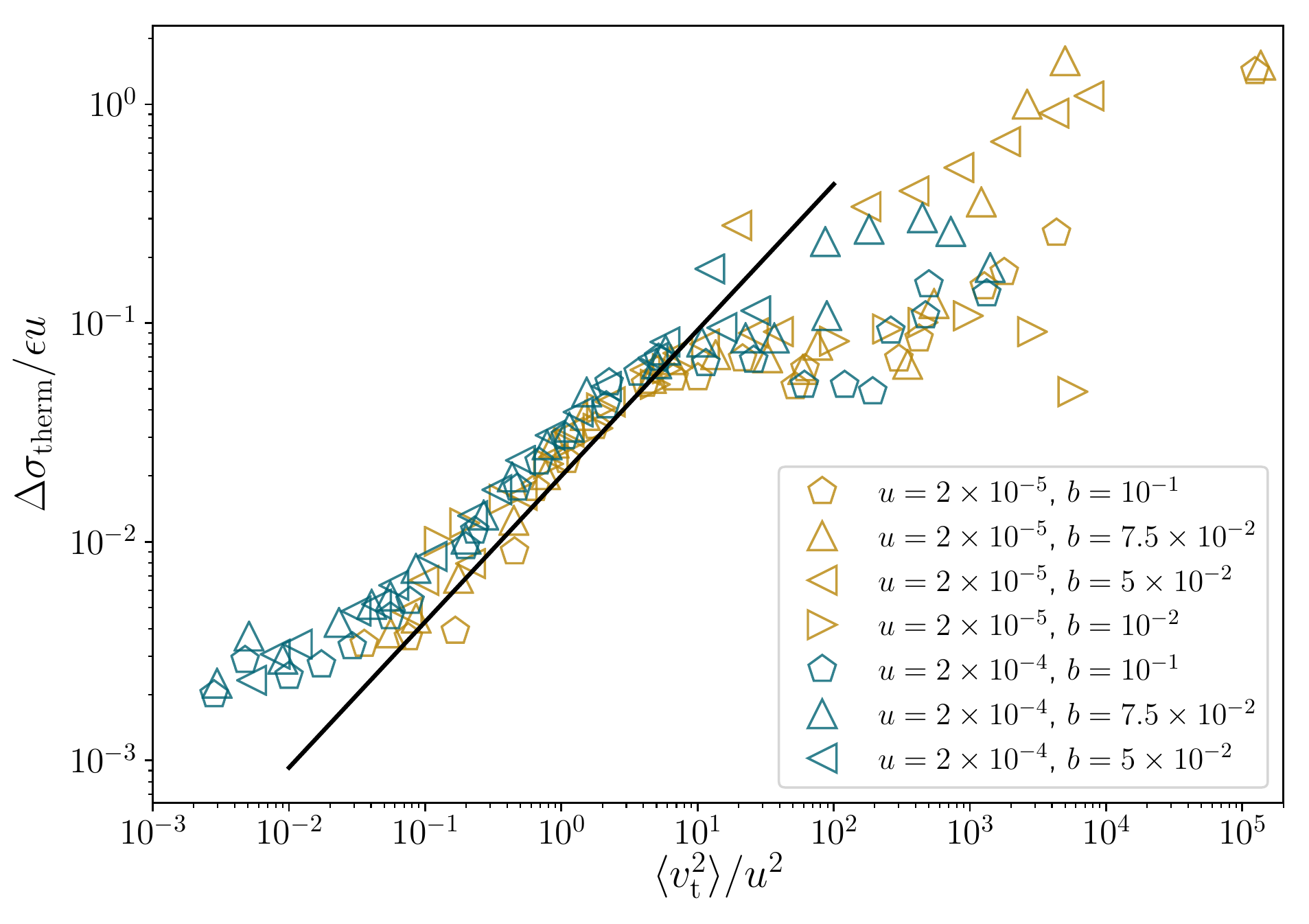}
  \caption{Thermal contribution of the stress
    $\Delta\sigma_{\rm therm}/\epsilon u$ vs. tangential velocities
    $v^2/u^2$. The line corresponds to
    $\Delta\sigma/u \propto (v^2/u^2)^{2/3}$. {Different colors
      correspond to different $u$.}  }
  \label{fig:dsigma}
\end{figure}

The energy barrier for a plastic rearrangement at a given strain
$\gamma_c$ is $\Delta E = k_BT_0(\gamma-\gamma_c)^{3/2}$, where
$\gamma$ is the currrent strain and $k_BT_0$ is the overall energy
scale of the process~\cite{PhysRevLett.105.266001}. We can take
$k_BT_0\sim \epsilon u^2$. Thermal activation is possible, when
$\Delta E\approx k_BT\sim mv^2$. Thus plastic yielding does, in
general, not occur at $\gamma_c$ (as it would at zero temperature) but
at reduced strains
$\Delta\gamma\sim (T/T_0)^{2/3}\sim (v^2/u^2)^{2/3}$. The associated
stress reduction is therefore
$\Delta\sigma_{\rm therm} =g\Delta\gamma$. The scaling is presented in
Fig.~\ref{fig:dsigma}~\footnote{A more detailed derivation of the
  underlying theory includes logarithmic corrections as presented in
  Refs.~\cite{PhysRevLett.105.266001,PhysRevLett.95.195501}.}. A
systematic deviation from the scaling behaviour at small velocities
(small strainrates) is due to the lack of precise value for the yield
stress. Much longer simulations at lower strainrates would be
necessary to overcome this limitation.

\subsubsection{Discontinuity}

\begin{figure}
  \includegraphics[width=0.49\textwidth]{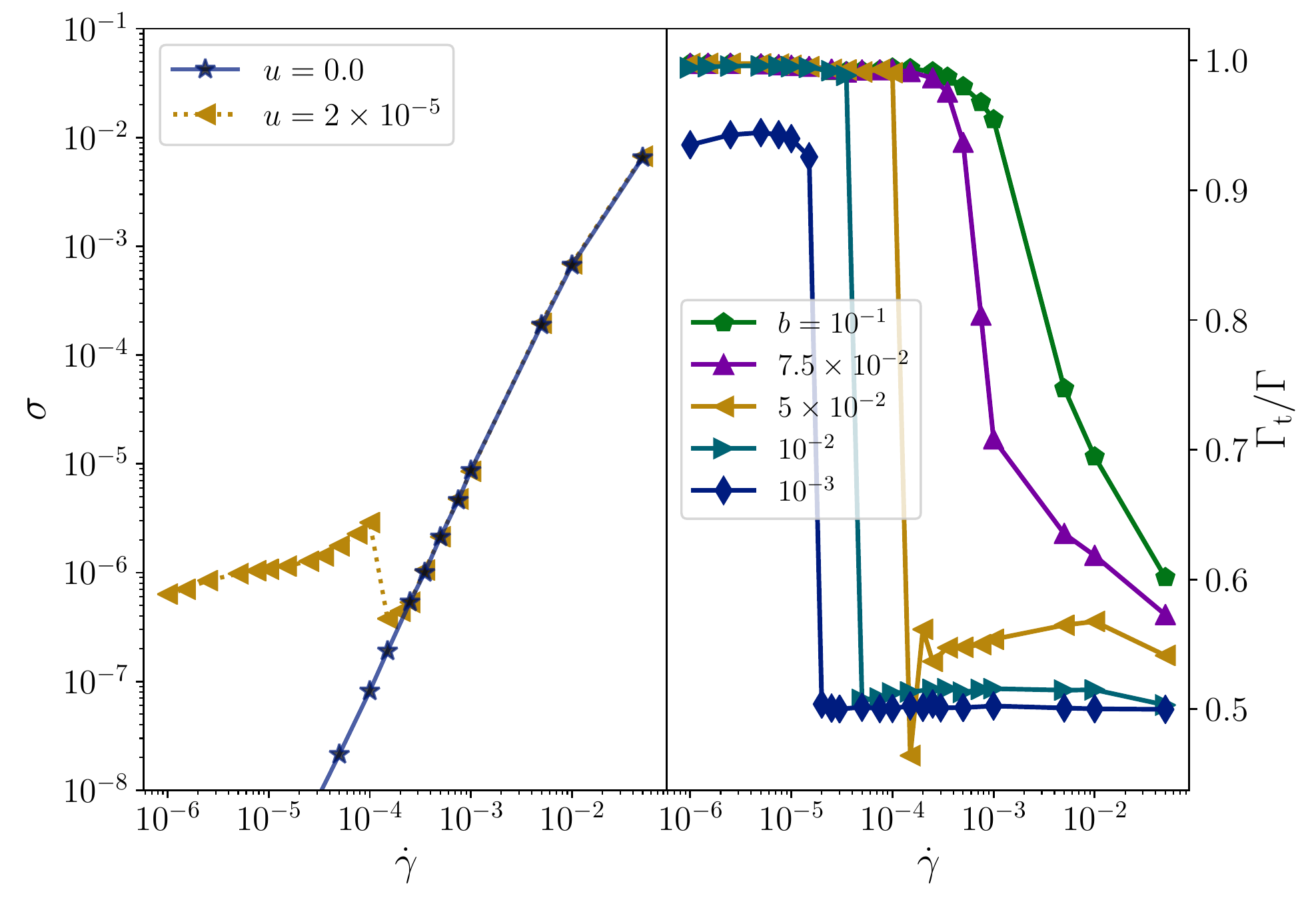}
  \caption{(Left) Flow curves of repulsive ($u=0.0$) and attractive
    ($u=2\times10^{-5}$) systems in the under-damped regime
    ($b=5\times10^{-2}$). The volume fraction is $\phi=0.75$. The
    repulsive curve exhibits the well-known Bagnold scaling,
    $\sigma \propto \dot\gamma^2$. (Right) The ratio of the tangential
    to total dissipation power in the under-damped regime. The
    attraction range and the volume fraction are $u=2\times10^{-5}$
    and $\phi=0.75$.}
  \label{fig:fc-underdamped-repatt-gamma}
\end{figure}

Now we turn to the discussion of the discontinuity in the flowcurves
of underdamped systems. Apparently, the yield-stress or viscous branch
of the flowcurve becomes unstable and the system jumps into an
inertial branch, characterized by Bagnold scaling,
$\sigma\sim \dot\gamma^2$. However, there does not seem to be a
continuous route between these two branches, unlike in the repulsive
systems reported in Ref.~\cite{FallPRL2010}. There, one finds a
continuous crossover into the inertial branch. Indeed, if we switch
off the attraction, we loose the discontinuity (see
Fig.~\ref{fig:fc-underdamped-repatt-gamma}, left
panel).

The crucial information is obtained when one splits the dissipated
energy rate $\Gamma$ (which determines the stress via
$\Gamma = \sigma\dot\gamma L^2$) into ``tangential'' and ``normal''
contributions $\Gamma = \Gamma_t +\Gamma_n$. As energy is only
dissipated in particle contacts, one can write
\begin{equation}
  \Gamma = \sum_{i=0}^{N} \vec{F}_{i,\text{diss.}}\cdot\vec{v}_{i}.
  \label{eq:gamma}
\end{equation}
In Eq.~(\ref{eq:gamma}),
$\vec{F}_{i,\text{diss.}}$ and $\vec{v}_i$ determine the damping force
(Eq.~(\ref{eq:fdiss})) exerted on and the velocity of the $i$th
particle. Splitting the particle velocities into components normal and
tangential with respect to the contact line of the two contacting
particles one obtains the associated contributions to the dissipated
power.

Fig.~\ref{fig:fc-underdamped-repatt-gamma} (right panel)
displays the ratio of tangential to total dissipation power
$\Gamma_t/\Gamma$. The discontinuity in the flow curve is visible as a
discontinuous drop of the fraction of dissipation in the tangential
channel. At small strainrates, as anticipated above, dissipation is
completely dominated by the tangential motion of particles around each
other. In the Bagnold branch at high strainrates, on the other hand,
dissipation is equally distributed in both channels, indicative of
random particle encounters in two-particle collisions.

Thus, we can conclude, that the origin of the discontinuity is
two-fold: first attractive particle interaction condense the particles
into a highly coordinated (yield-stress) fluid state, that involves an
overwhelming tangential contribution to energy dissipation. This fluid
comes to rest at the yield stress. When, by increasing strainrates,
particle momenta become too large, the cohesive force is marginalized
and the network is destroyed resulting in a gas-like state with
two-particle collisions and equi-partition between normal and
tangential energy dissipation.

These findings allow for a comparison with the phenomenon of
friction-induced shear thickening~\cite{ClausPRE2013}. In that
scenario it is the solid-solid friction between particles that allows
for a tangential channel of energy dissipation. This incurs the
coexistence of two branches in the flowcurves, a yield-stress branch
and a flowing state, that can either be viscous or Bagnold in
nature. When the stress reaches a certain threshold the flowing state
becomes metastable, followed by a discontinuous transition into the
yield-stress branch, which has a much higher stress. This gives rise
to the phenomenon of \emph{discontinuous shear thickening}.

Here, the presence of a tangential dissipation channel also gives rise
to the two states with high and low $\Gamma_t$,
respectively. Interestingly, however the stabilities of the two
branches are reversed as compared to the frictional scenario: the
yield stress branch is stable at small strainrates, while the Bagnold
fluid is stable at high strainrates. Thus, in general the stress
decreases discontinuously, and one can speak of \emph{discontinuous
  shear thinning}. One has to be careful, however. There is one
instance in Fig.~\ref{fig:fc-z-u2e-5}, where the stress does not
decrease but increase. This happens at the smallest available damping
$b=10^{-3}$. The reason for this inversion is the opposite dependence
on $b$ of the two branches. While the yield stress branch decreases,
the Bagnold branch increases upon decreasing $b$.

\begin{figure}
 \includegraphics[width=0.5\textwidth]{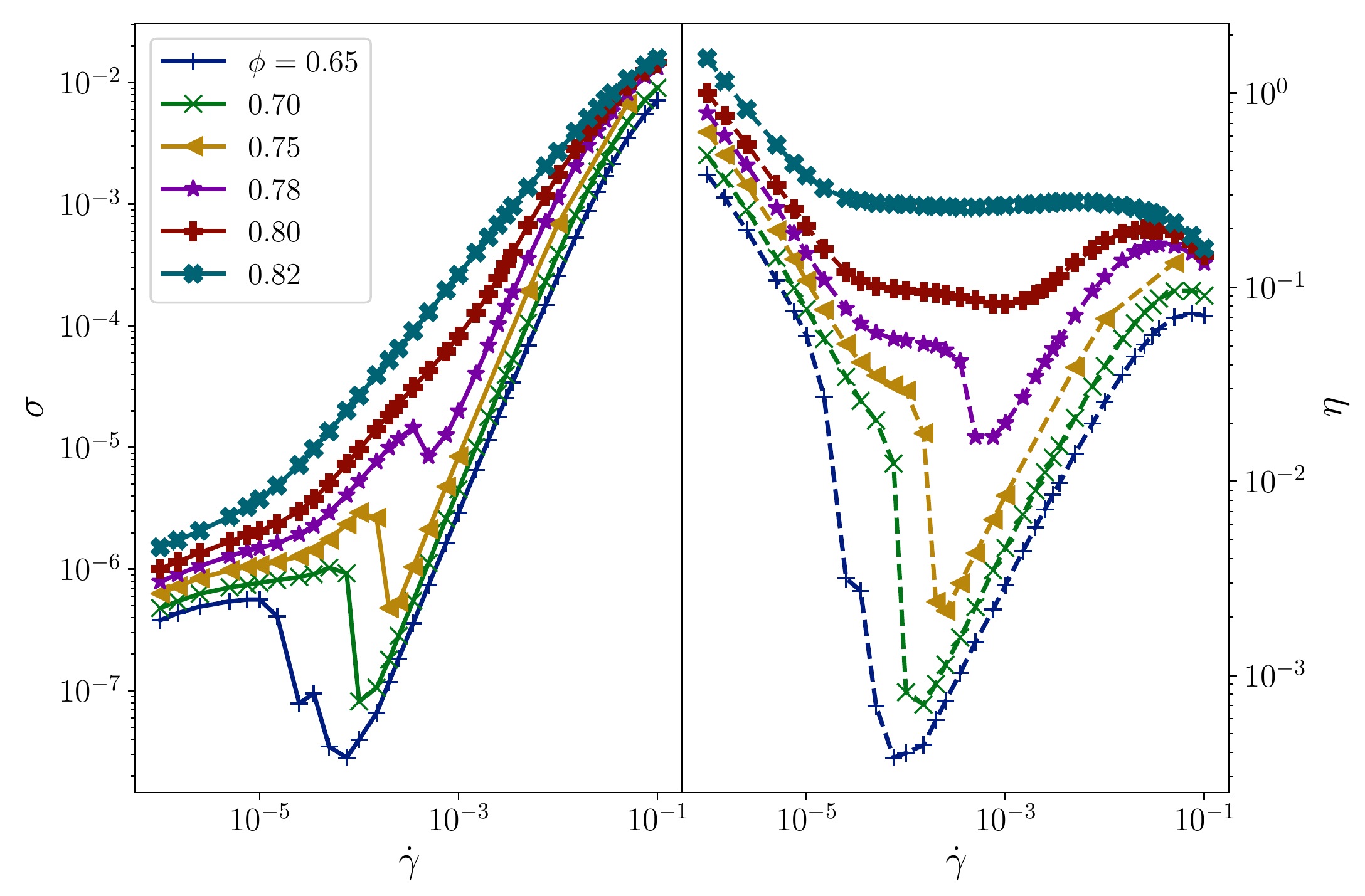}
  \caption{(Left) Flow curves, i.e. stress $\sigma$ vs. strainrate
    $\dot\gamma$, at $u=2\times10^{-5}$, $b=5\times10^{-2}$ and
    different $\phi$. (Right) Corresponding data for viscosity, $\eta({\dot{\gamma}})$.}
  \label{fig:cd-diff-phi}
\end{figure}

Pursuing the analogy with the frictional scenario further, we test for
the dependence on volume fraction $\phi$. 
In Fig.~\ref{fig:cd-diff-phi},
we show the flow curve as well as the corresponding viscosity, for one such set
of $u$ and $b$, with changing $\phi$.
The discontinuous shear thinning, from the yield stress branch to the inertial branch,
evolves out of a continuous shear thinning, quite similar to the
frictional scenario. Again, the trend is reversed, however. With
friction, it is the increase of volume fraction that triggers the
transition from continuous to discontinuous shear
thickening~\cite{0295-5075-115-5-54006}. Here, it is the decrease of
volume fraction. This follows from the reversed stability in terms of
strainrate and the fact that the yield-stress branch is more stable at
higher volume-fraction.
In the frictional system, because of the properties of the frictional
interaction, it is only at high pressures that particles can condense
into network-like structures. Then it is possible that the frictional
dissipation is enhanced beyond what one would expect from simple
two-particle collisions in a granular gas. Pressure increases both
with volume-fraction and strainrate. On the other hand, adhesion
strength is independent of strainrate and faster motion thus weakens
the network and triggers a transition into the fluid.

\subsubsection{Shear bands}

With the phenomenon of discontinuous shear thinning we open up the
possibility for flow instabilities.

\begin{figure}
  \includegraphics[width=0.4\textwidth]{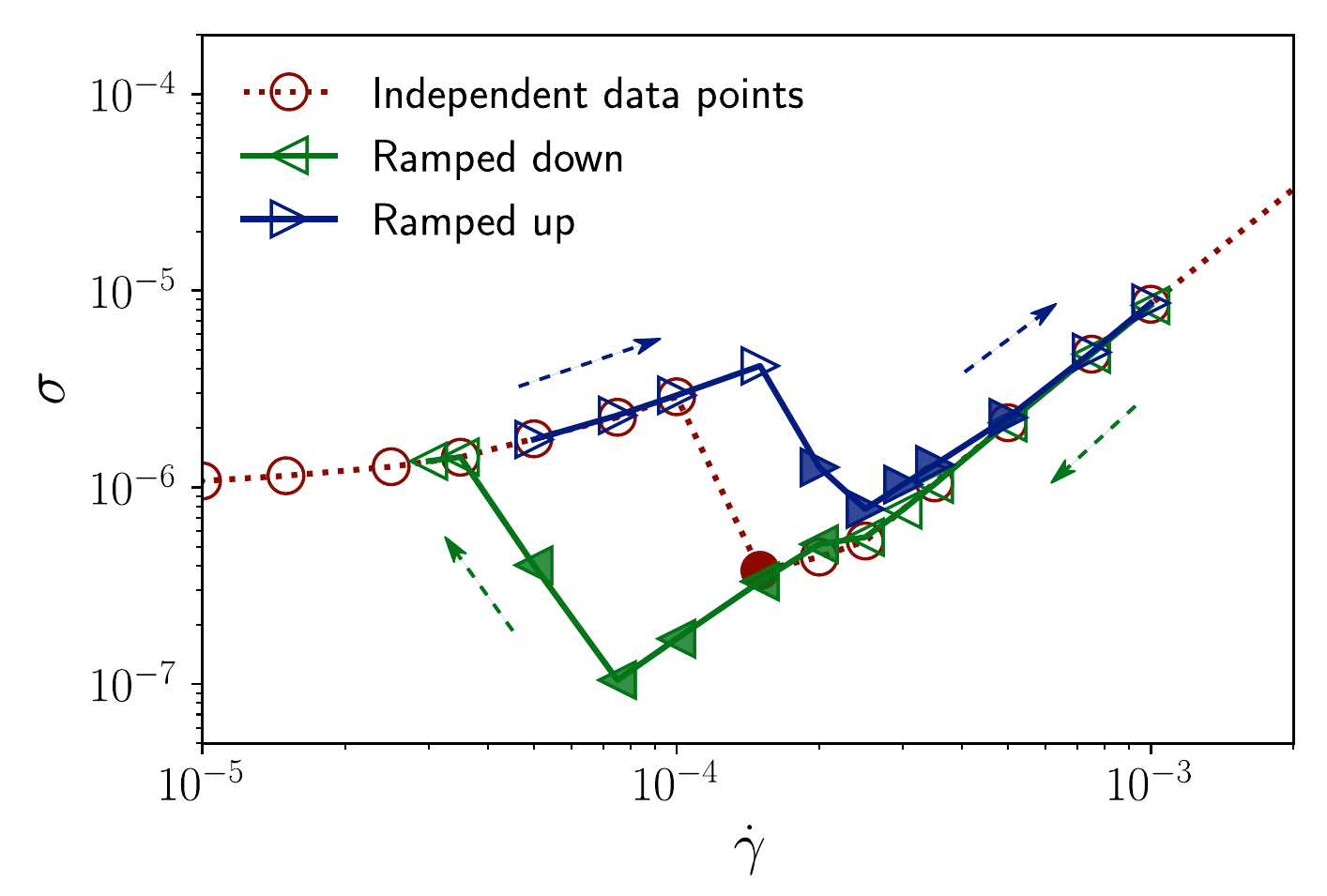}
  \caption{Flowcurve for a strainrate ramp. Hysteresis and shear
	bands for a system with $u=2\times10^{-5}$, $\phi=0.75$ and $b=5\times10^{-2}$.
	Filled symbols indicate where shear bands are observed. }
  \label{fig:flowcurve-ramp}
\end{figure}

%\xx{
In the frictional scenario of discontinuous shear-thickening
  unsteady chaotic flow
  \cite{PhysRevE.93.030901,doi:10.1122/1.4953814} and vorticity
  banding has been observed~\cite{PhysRevLett.121.108003}. On the
  other hand, a decreasing flowcurve is prone to the formation of
  shear bands in the gradient direction%}. 
Long simulations in large
enough systems are necessary to make these instabilities observable.
Indeed, we can identify shear bands in the vicinity of the
discontinuity when we perform ramping simulations. In these simulation
we slowly ramp up the strainrate after a strain of
$\gamma_{\rm tot}=40$, until we reach high enough strainrates and the
ramp is reversed. The associated flowcurve encircling the
discontinuity is depicted in Fig.~\ref{fig:flowcurve-ramp}.
Next to an extended hysteresis loop we find evidence of shear banded
states. These are highlighted by filled symbols in the
figure. Interestingly, the average stress in these shear-banded states
does not differ much from the value in the homogeneous state (it is
somewhat larger). Also, there is no indication of a stress
plateau. Such a plateau would be indicative of a scenario where the
two coexisting bands represent states (at equal stress) from an
underlying non-banded local flowcurve~\cite{DhontPRE99}.

\begin{figure}
  \includegraphics[width=0.5\textwidth]{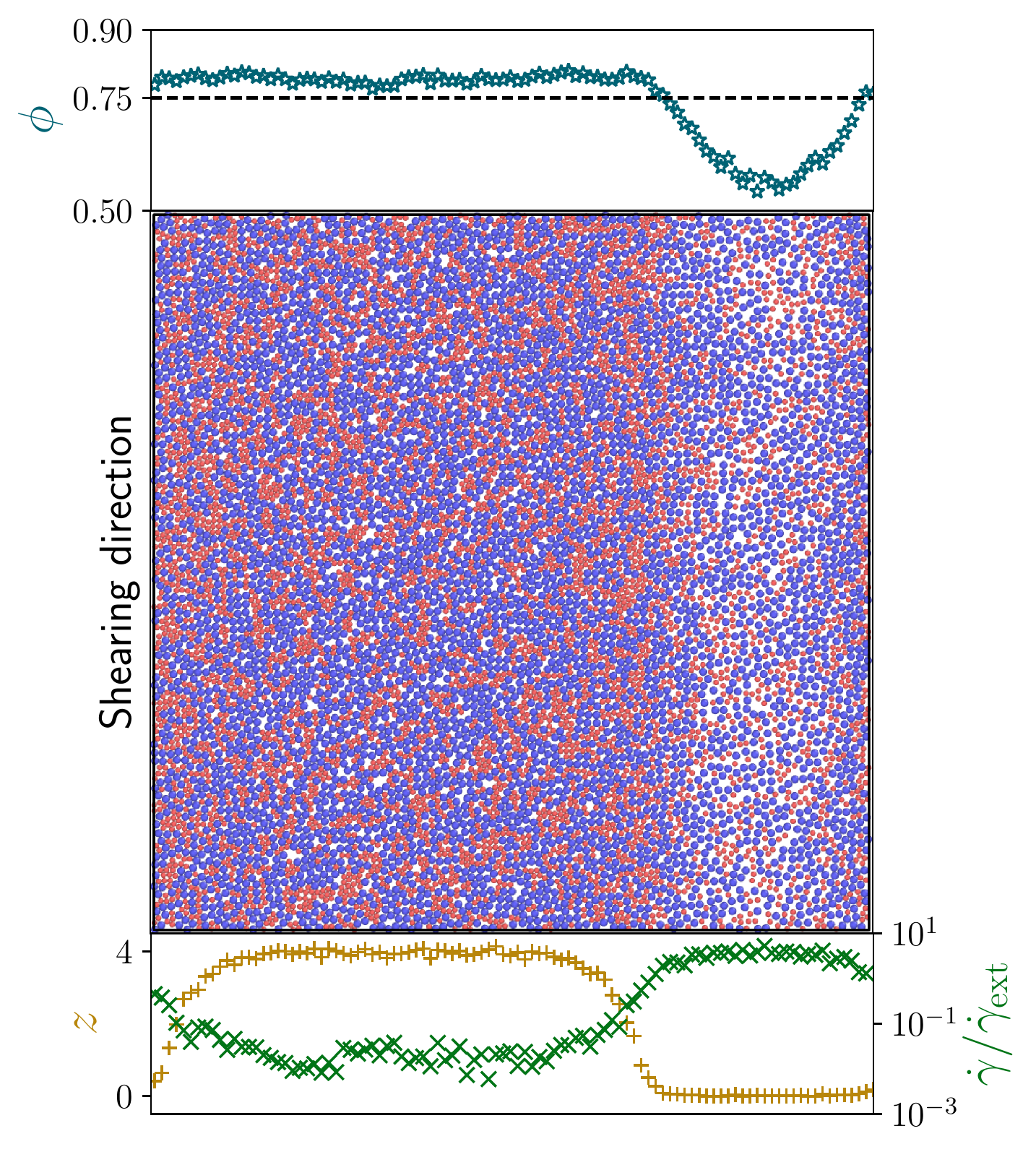}
  \caption{Snapshot of a shear-banded system with $N=10^4$,
    $b=5\times10^{-2}$, $u=2\times10^{-5}$, $\phi=0.75$ and imposed
    strainrate $\dot\gamma_{\rm ext}=2\times10^{-4}$. Panels at top
    and bottom highlight the local values of volume fraction $\phi$ as
    well as connectivity $z$ and strainrate
    $\dot\gamma/\dot\gamma_{\rm ext}$. These are obtained by
    performing averages along slices in shearing direction.}
  \label{fig:snap-cd-underdamped}
\end{figure}

A snapshot of a shear-banded state makes clear what is happening
(Fig.~\ref{fig:snap-cd-underdamped}). Next to the strainrate, the
volume fraction as well as the connectivity vary between the bands.
Volume fraction and connectivity are substantially reduced in the
inertial band, indicating a dilute granular gas state. In the remaining system
the connectivity is close to the threshold value of $z_{\rm iso}=4$,
which would indicate the possibility of a solid state. It might
therefore represent a quasi-solid band at a slightly elevated
volume-fraction $\phi>0.75$. Interestingly, the interface between the
two bands seems to have an even higher density, even though the
connectivity is markedly reduced and interpolates from the high value
of the solid to the small value of the gas. Such an anti-correlation
between connectivity and density is also observable in our previous
work \cite{IraniPRL2014,IraniPRE2016}, where the tangential channel of
energy dissipation is absent. In that system, however, the gas-like
band is absent and the solid band coexists with what here might be just an
interface. We also note that such changes in local density, during
shearband formation, has also been observed in experiments
involving discontinuous shear-thickening \cite{PhysRevLett.114.098301}.

Larger systems are necessary to study these questions in more
detail. It might also be interesting to add a third spatial dimension
to study the possibility of vorticity banding in the presence of
discontinuities in the flowcurves~\cite{PhysRevLett.121.108003}.

\section{Conclusion}

The system is quite similar to the one studied by Nicolas {\it et al.}
\cite{NicolasPRL2016} with two important differences. The first relates to the presence, in
the current work, of a tangential channel of energy dissipation. The
second to the type of adhesion forces.
Nicolas {\it et al.} use a standard Lennard-Jones interaction,
where the range of attraction is on the order of the diameter of the
repulsive core. In our system repulsive core (particle diameter $d=1$)
and attraction range $ud$ are scale separated and $u\ll
1$. Thus, attractive forces are really only active when particles are
near contact.

The resulting solid is therefore very fragile in the sense that the
number of interactions (``contacts'') is just slightly above the
minimum isostatic value, $\delta\!z = z-z_{\rm iso}\ll 1$.  This
allows us on the one hand to derive a scaling expression for the yield
stress $\sigma_y\sim u^{1/2}\delta\!z^{3/2}$. On the other hand, the
yield stress is very small and not accessible in our work. Instead we
observe an extended Herschel-Bulkley (HB) regime
$\sigma\sim \dot\gamma^\nu$ with an exponent $\nu\approx 0.45$ quite
similar to other studies (Nicolas {\it et al.} have $\nu=0.5$). We
tried to also relate this exponent to the underlying isostatic
structure but so far without success.

In underdamped systems a weakening effect sets in and the stress is
reduced below the HB branch. This is due to an effective temperature
that originates in enhanced velocity fluctuations in weakly damped
systems. The resulting $T^{2/3}$ scaling of the stress follows the
theory proposed in
Refs. \cite{PhysRevLett.95.195501,PhysRevLett.105.266001}.
The same scaling is observed in Nicolas {\it et al.} however, the
scaling of the shear-induced temperature $T$ itself is different in
that work. There, the energy balance argument Eq.~(\ref{eq:en_balance}) is used
to derive (the equivalent of our notation)
\begin{equation}
    T\sim m\dot\gamma^2\text{Wi}^{-1}
\end{equation}
which is valid as long as the stress scale is set just by attractive
forces $\sigma\sim \epsilon u$. Here, however, it is far away from the
yield point deep in the flowing region, where the weakening effect
sets in. The relevant stress scale therefore is the HB result,
$\sigma\sim \epsilon u\text{Wi}^\nu$ and
$T\sim m\dot\gamma^2\text{Wi}^{\nu-1}$. This is also why for weak
inertial effects the stress does not decrease as a function of
strainrate, as in Ref.\cite{NicolasPRL2016}. Rather, the stress
reduction $\Delta\sigma_{\rm th}$ superimposes on the HB law
$\sigma\sim \epsilon u \text{Wi}^\nu$, which in effect leads to a
weaker increase or to an intermediate plateau in stress (see
Fig.~\ref{fig:fc-z-u2e-5}). Only for the weakest damping do we
actually observe a negative slope in the flowcurve.

Finally, no shear bands are observed in Nicolas {\it et al.}. Here, we
do observe a shear banding instability, albeit the scenario is quite
complex and associated with a discontinuity in the flowcurve which
does not directly follow from the weakening effect just described.

Rather, we set up an analogy with frictional systems, where the
phenomenon of {\it discontinuous shear-thickening} can be interpreted
as a coexistence (spatial or temporal) of two disconnected branches of
the flowcurve. Due to the nature of the frictional interaction, the
fluid branch is stable at low volume-fractions and small
strainrates. Increasing the strainrates the system jumps to the HB
branch, which is at higher stress.  In the adhesive system discussed
here, the same two branches are present but the roles and stabilies
are reversed. At small densities and strainrates the solid HB branch
is stable -- the particles condense into a network- or gel-like
inhomogeneous structure. Increasing the strainrate the structure of
the network is weakened and the system jumps into the fluid branch,
which is at lower stress. Thus the system undergoes \emph{discontinuos
  shear thinning}. The crucial ingredient to observe these
discontinuities in both cases is a tangential channel of energy
dissipation, i.e. due to velocities directed tangentially to the
particle surface. It will be interesting to see if these effects can
be generalized to other dissipation models, for example lubrication
forces, which also have normal and tangential modes. Future work
should also study Brownian systems to address the interplay between
real and shear-induced temperature.

\begin{acknowledgments}
  We acknowledge financial support by the German Science Foundation
  via the Emmy Noether program (He 6322/1-1).
\end{acknowledgments}

%\bibliographystyle{unsrt}
%\bibliography{e2}


\end{document}